\documentstyle[aps,epsfig,multicol]{revtex}

\newcommand{\smx}{Sm$_{2-x}$Ce$_x$CuO$_{4-\delta}$}

\newcommand{\ndx}{Nd$_{2-x}$Ce$_x$CuO$_{4-\delta}$}
\newcommand{\lax}{La$_{2-x}$M$_x$CuO$_4$}
\newcommand{\lnx}{Ln$_{2-x}$M$_x$CuO$_4$}
\newcommand{\scco}{Sm$_{1.85}$Ce$_{0.15}$CuO$_{4-\delta}$}
\newcommand{\ncco}{Nd$_{1.85}$Ce$_{0.15}$CuO$_{4-\delta}$}
\newcommand{\psm}{Sm$_2$CuO$_4$}

\newcommand{\pla}{La$_2$CuO$_4$}
\newcommand{\pln}{Ln$_2$CuO$_4$}
\newcommand{\degree}{$^\circ$C}
\newcommand{\be}{\begin{equation}}
\newcommand{\ee}{\end{equation}}
\newcommand{\cp}{$C$$\rm{_p}$}
\newcommand{\Hs}{{\bf H} $\perp$ {\bf c}}
\newcommand{\Hp}{{\bf H} $\parallel$ {\bf c}}
\newcommand{\Tc}{$T$$\rm{_c}$}
\newcommand{\TN}{$T$$\rm{_N}$}

\begin{document}
\draft
\title{Electronic and Magnetic Properties of Electron-doped Superconductor, \scco}

\author{B. K. Cho}
\address{Department of Materials Science and Engineering,
         K-JIST, Kwangju 500-712, Korea}
\author{Jae Hoon Kim, Young Jin Kim}
\address{Institute of Physics and Applied Physics, Yonsei University,
         Seoul 120-749, Korea}
\author{Beom-hoan O}
\address{Department of Electronic Materials and Devices Engineering,
         Inha University, Inchon 402-751, Korea}
\author{J. S. Kim, G. R. Stewart}
\address{Department of Physics, University of Florida, Gainesville,
         Florida 32611-8440, USA}

\date{submitted to Phys. Rev. B. July 18, 2000. Revised: August 13, 2000}
\maketitle

\begin{abstract}
Temperature-dependent magnetization ($M$($T$)) and specific heat
(\cp($T$)) measurements were carried out on single crystal \scco\
($T_c$ = 16.5 K).
The magnetic anisotropy in the static susceptibility,
$\chi \equiv$ $M/H$, is apparent not only in its magnitude
but also in its temperature dependence,
with $\chi_{\perp}$ for \Hs\ larger than $\chi_{\parallel}$ for \Hp.
For both field orientations, $\chi$ does not follow the Curie-Weiss
behavior due to the small energy gap of the $J$ = 7/2 multiplet
above the $J$ = 5/2 ground-state multiplet.
However, with increasing temperature,
$\chi_{\parallel}$($T$) exhibits a broad minimum near 100 K and
then a slow increase  while $\chi_{\perp}$($T$)
shows a monotonic decrease.
A sharp peak in \cp($T$) at 4.7 K manifests an antiferromagnetic
ordering.
The electronic contribution, $\gamma$, to \cp($T$) is estimated
to be $\gamma$ = 103.2 (7) mJ/mole${\cdot}$Sm${\cdot}$K$^2$.
The entropy associated with the magnetic ordering is much smaller
than $R$ln2, where $R$ is the gas constant,
which is usually expected for the doublet ground state of Sm$^{+3}$.
The unusual magnetic and electronic properties evident in $M$($T$)
and \cp($T$) are probably due to a strong anisotropic interaction
between conduction electrons and localized electrons
at Sm$^{+3}$ sites.
\end{abstract}


\begin{multicols}{2}

\section{introduction}

After the discovery of high-temperature superconductivity
in copper-oxide compounds \cite{Bednorz},
a new class of superconducting compounds was found
with the formula \lnx\ where Ln stands
for Pr, Nd, Sm, and Eu, and M for Ce and Th \cite{Tokura}.
These compounds have become the subjects of an intense study
due to their peculiar physical properties,
which are different from the high-temperature
superconductors of cuprates.
The \pln\ parent compounds crystallize in a tetragonal
``\,$T^{\prime}$-phase\,'' structure containing CuO$_2$ planes
in which the copper ions are surrounded
by a square planar arrangement of oxygen ions,
in contrast to the \pla\ parent compound
which forms an orthorhombic ``\,$T$-phase\,'' structure
at low temperature (below $\approx$ 500 K)
containing CuO$_2$ planes in which copper ions
are surrounded by an octahedral arrangement of oxygen ions.
\lnx\ (M = Ce or Th) compounds have electrons as a charge carrier
in forming superconducting electron pairs,
contrast to the related \lax\ (M = Sr or Ba) compounds
containing holes as a charge carrier.
The electron-doped compounds have the pressure dependence
of \Tc\ variation with negative $d$ln\Tc/$dP$,
where $P$ is pressure, while the hole-doped ones have positive
$d$ln\Tc/$dP$ \cite{Murayama}.
Antiferromagnetic (AFM) ordering of the rare-earth ions
in \pln\ has been found for Ln = Nd (\TN\ $\approx$ 1.7 K),
Sm (\TN\ $\approx$ 5.9 K), and Gd (\TN\ $\approx$ 6.6 K)
while no magnetic ordering for Ln = Pr and Eu.
The AFM ordering temperatures for Ln = Nd and Sm are lowered
by substituting electron donor element
(Ce$^{+4}$ or Th$^{+4}$ ions) for Ln$^{+3}$ ions.
The superconductivity appears at the narrow range of electron doping
near $x = 0.15$ with \Tc\ $>$ \TN\ and co-exists
with the AFM state below \TN.
The AFM transition nature is studied in terms of magnetization
and specific heat measurements \cite{Ghamaty,Seaman}.
The entropy estimation associated magnetic ordering in \psm\ confirms
the doublet ground state,
expected by crystalline electric field splitting \cite{Nekvasil}.
However, the electronic contribution to specific heat,
$\gamma$ $\approx$ 82 mJ/mole${\cdot}$Sm${\cdot}$K$^2$,
in \psm\ is found to be much larger than the ones of other compounds.
The large value of $\gamma$ is suspected to be due to the existence
of magnetic correlation much above the \TN,
making the evaluation of $\gamma$ uncertain,
but not understood clearly yet.

For superconductivity, experimental determination of the
order-parameter symmetry of $n$-type cuprate superconductors is
critical in establishing an unified understanding of the mechanism
of superconducting pairing in cuprates. Recent experiments suggest
that the dominant symmetry of the order parameter is of $d$-wave
type \cite{Kokales,Prozorov,Tsuei}. In addition, the role of
rare-earth magnetic moment interacting with $d$-wave superconducting
system of electrons opens up a new area of theoretical and
experimental studies. So far only \ndx\ was extensively studied in
which relatively weak moments strongly influence the temperature
dependence of the penetration depth \cite{Kokales,Prozorov}, which
helps to identify the order-parameter symmetry. It is interesting
to investigate the electronic and magnetic properties in normal
state of electron doped \scco\ by specific heat measurements.
Temperature dependent magnetization in the normal state is also
necessary to study the specific heat data because the crystalline
electric field effects significantly affect the magnetic
properties of Sm$^{+3}$ ion, causing magnetic anisotropy both in
magnitude and temperature dependence of magnetization. In this
paper, the specific heat and magnetization data are represented to
study the normal state properties of superconducting \scco\
compounds.

\section{experimetal details}
Superconducting single crystals of \scco\
have been grown by a flux-based technique.
A batch of about 40 g is prepared
by mixing and grinding powders of
Sm$_2$O$_3$ (99.9\%), CeO$_2$ (99.99\%), and CuO (99.99\%)
in the molar ratio of (2-$x$):(2$x$):(7.2$\sim$13.4), respectively.
The powders were pre-baked
at 800$-$950 \degree\ (for Sm$_2$O$_3$ and CeO$_2$)
or at 400$-$600 \degree\ (for CuO) to remove some volatile impurities.
The mixed batch needs to be sintered at 900 \degree\
and ground several times.
It was soaked at 1000 \degree\ for 10$-$20 hours
and heated to 1210 \degree\ in air (300 \degree/h).
After a short soak for 1$-$3 h,
the temperature was lowered to 1000 \degree\ at a rate of 5$-$12 \degree/h,
and then to room temperature.
As-grown crystals with typical size of
$\sim1.5\times1\times0.03 \mbox{ mm}^3$
were synthesized by this procedure.
Superconductivity was induced by annealing and quenching in inert gas;
the initial raising rate of temperature was 5$-$10 \degree/min
(300$-$600 \degree/h),
and the soak time at 880 \degree\ was 16 hours.
The quenching needs to be done within 30 minutes
to preserve the high-temperature structure.

The as grown single crystals of \scco\ are confirmed to be
of the single phase of the \psm\ structure
by measurements of powder x-ray diffraction of pulverized single crystals.
The impurity phases of Cu$_2$O and Sm$_2$O$_3$,
which are often found in polycrystalline samples,
are not detected in the diffraction pattern.
Temperature dependent static magnetization was measured by
using a 7-Tesla Quantum Design superconducting quantum interference
device magnetometer(SQUID).
The field cooled (FCW) and zero-field-cooled (ZFC) data
in the superconducting state
were obtained on warming after the magnet was quenched.
The specific heat measurements down to 1.2 K were made
on the grown single crystal,
using a time constant method (relaxation method) technically described
in detail elsewhere \cite{Stewart1}.

\section{results and discussion}

\begin{figure}
 \centerline{\psfig{figure=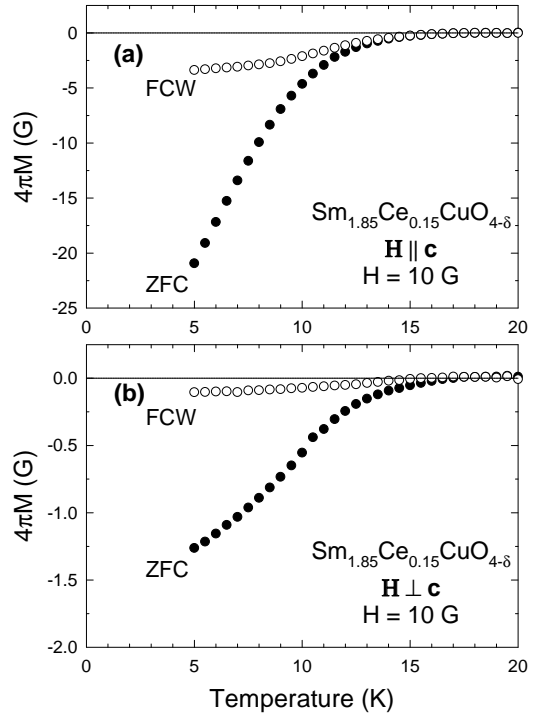,width=7cm,clip}}
 \caption{Superconducting state volume magnetization $M$ in an applied
  field $H$ = 10 G versus temperature of single crystal \scco\ for 
  (a) \Hp\ and
  (b) \Hs: zero-field-cooled (ZFC) (dark circles) and field-cooled (FCW)
  (open circles) data taken on warming as shown.}
 \label{fig1}
\end{figure}

The magnetization versus temperature ($M$($T$)) data in
Fig.~\ref{fig1}(a) and \ref{fig1}(b) show the flux expulsion (FCW) and
magnetic shielding (ZFC) effects for \Hp\ and \Hs\ in a  \scco\
crystal for an external magnetic field $H$ = 10 G, respectively. The
plots show typical superconducting diamagnetic signal for both
field orientations, indicating bulk superconductivity. The much
higher values of $M$($T$) for \Hp\ than \Hs\ is due to the
demagnetization field inside the sample, which is not corrected
for actual real field for the measurements. The superconducting
transition temperature, \Tc, is found to be 16.5 K, the temperature
at which more than 1\% of superconducting volume fraction appears.
It is noted that the superconductivity in \ndx\ appears only both
in the very limited Ce concentration range of $x$ $\approx$ 0.15
and in the reduced oxygen content of $\delta$ $\approx$ 0.07
\cite{Ghamaty}. The superconducting properties of \scco\ single
crystal are quite similar to the ones of bulk \ndx\ samples,
indicating the apparent oxygen deficiency in our sample. In
addition, the observed \Tc\ $\approx$ 16.5 K and the broad
superconducting transition in the magnetization in low magnetic
fields is often found in the bulk \scco\ samples, due to the
partial occupancy of apical oxygen in $T$-phase structure
\cite{Radaelli}. It should be noted that, recently, the microwave
surface resistance measurement, which depends {\it neither} on
electric percolation nature {\it nor} on magnetic shielding
current, shows that the real \Tc, clearly higher than the \Tc\
determined above, exists without measurable bulk Meissner effect
in \smx\ compounds \cite{Blackstead}. So the \Tc, which is
determined in this study, is believed to be lower bound of real \Tc.

\begin{figure}
 \centerline{\psfig{figure=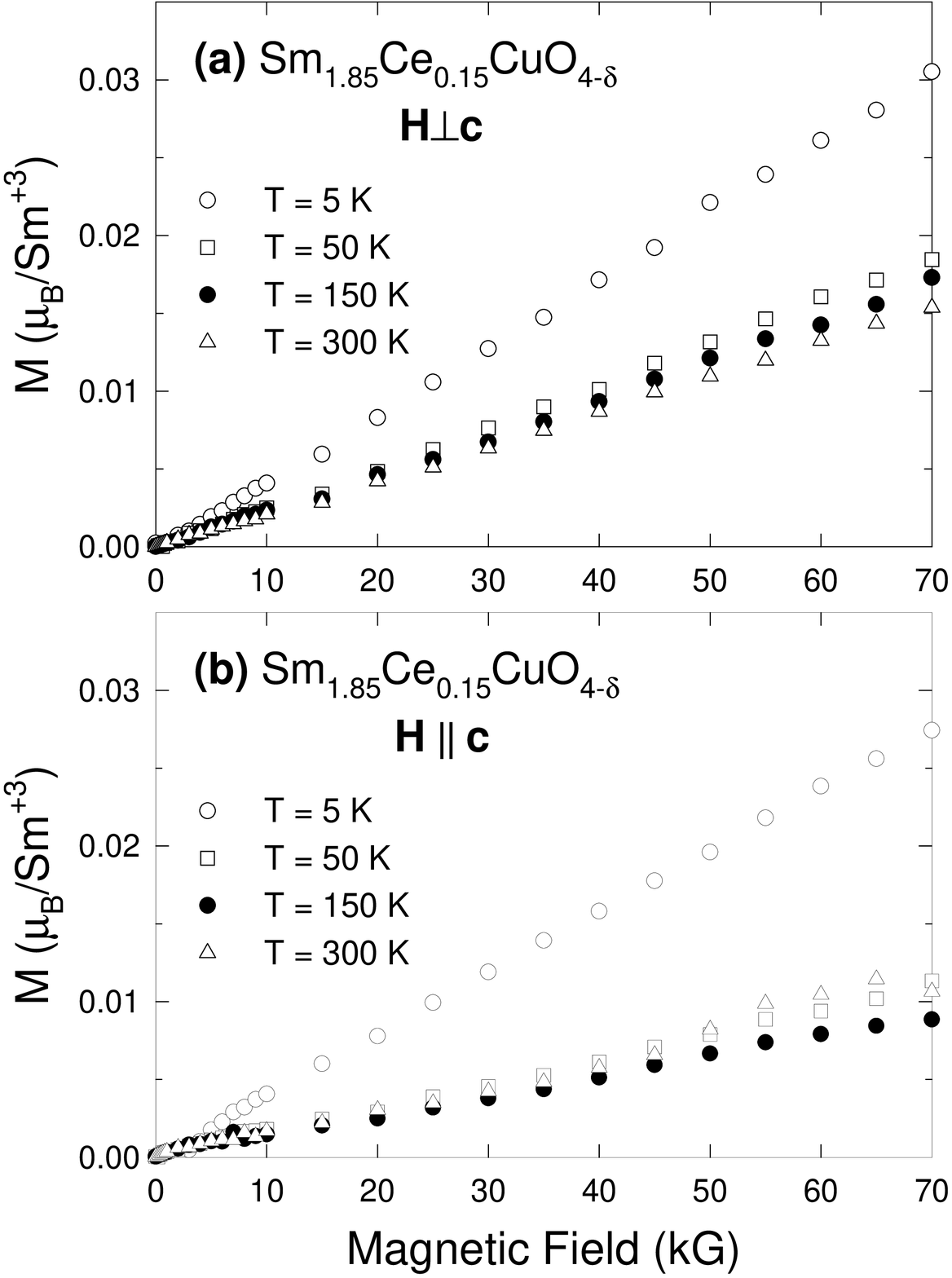,width=7cm,clip}}
 \caption{Magnetization $M$ versus applied magnetic field $H$
   of single crystal \scco\ at the indicated temperature:
   (a) \Hs\ and (b) \Hp.}
 \label{fig2}
\end{figure}

Typical $M$($H$) isotherm data for \scco\ are shown
in Fig.~\ref{fig2}(a) for \Hs\
and \ref{fig2}(b) for \Hp\
at several different temperatures for 0 G $\le H \le$ 70 kG.
For both field orientations, the magnetization is linear
in the whole applied field range for temperature
above 50 K and at 5 K in the $H >$ 10 kG
below which superconducting signals appear.
It is noted that the nonlinear behavior of magnetization,
leading to a saturation of the Sm$^{+3}$ magnetic moments,
is not observed even at $T$ = 5 K and $H$ = 70 kG.
The magnetic moment at this temperature and field is found to be
0.031 $\mu_{\rm B}$/Sm$^{+3}$
and 0.027 $\mu_{\rm B}$/Sm$^{+3}$ for \Hs\ and \Hp, respectively.
Those values are much smaller than
the theoretically expected value of 0.845 $\mu_{\rm B}$/Sm$^{+3}$
for Hund's isolated Sm$^{+3}$ ion, $^6$H$_{5/2}$.

\begin{figure}[t]
 \centerline{\psfig{figure=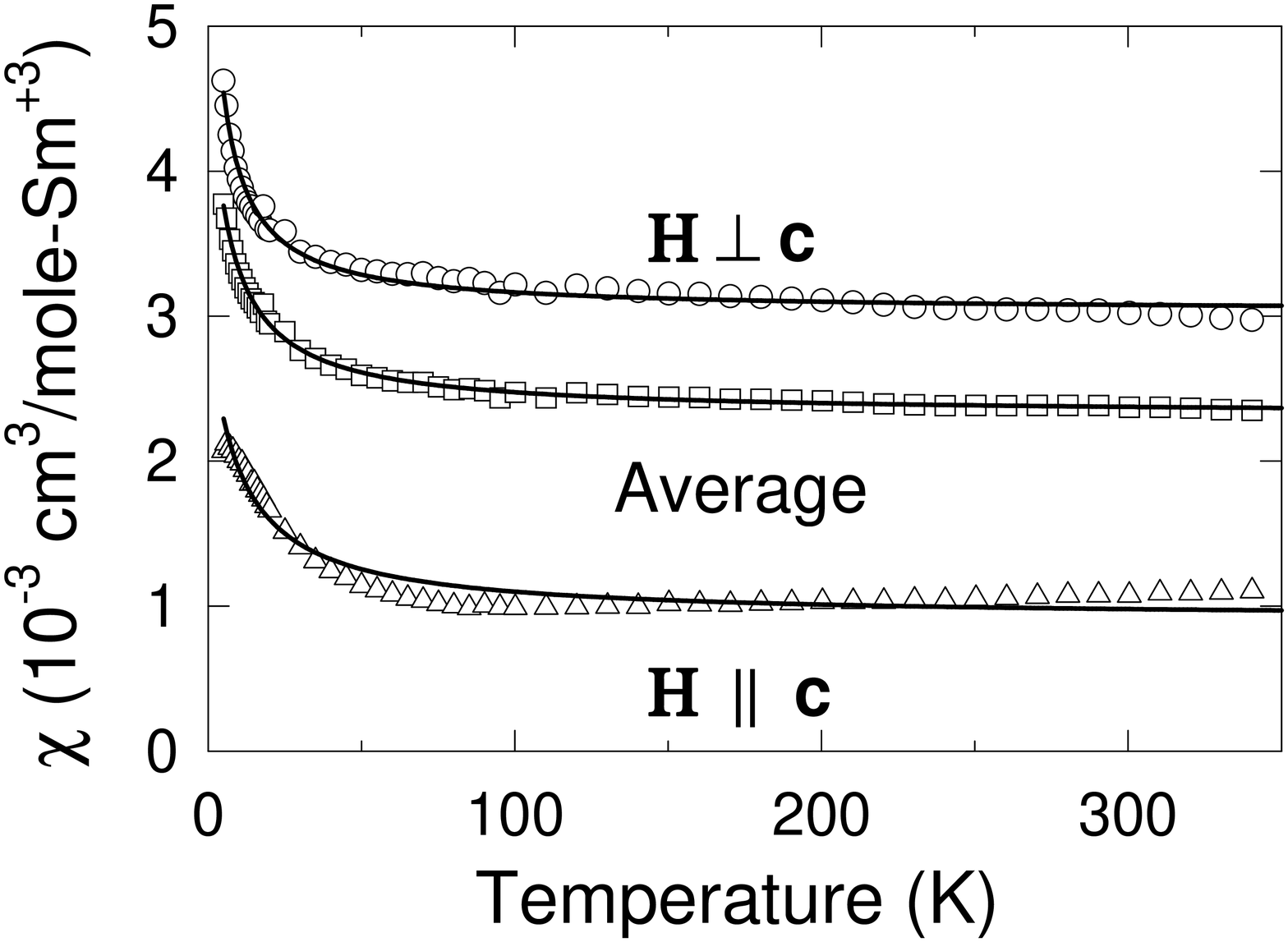,width=7cm,clip}}
\caption{Anisotropic magnetic susceptibility $\chi$
  versus temperature $T$ of single crystal \scco\ 
  for 5 K $\le T \le$ 350 K
  for \Hs, \Hp, and powder average.
  Fits to Equation (\ref{chi}) in the text are shown
  by the solid curves for each field orientations.}
\label{fig3}
\end{figure}

Fig.~\ref{fig3} shows the temperature-dependent
magnetic susceptibility, $\chi$($T$), for \scco\ with $H$ = 5 kG
perpendicular and parallel to $c$-axis and their
powder average for 5 K $\le T \le$ 350 K.
The large anisotropy in $\chi$($T$) between \Hs\ and \Hp is quite
clear and  the temperature dependence for both field orientations
clearly deviate from the typical Curie-Weiss behavior.
In addition, the temperature dependences of $\chi$($T$)
for both field orientations is also significantly different:
with increasing temperature,
$\chi_{\parallel}$($T$) for \Hp\ shows a broad local minimum
around 100 K and a slow increase whereas $\chi_{\perp}$($T$) for \Hs\
shows a monotonic decrease.
The similar $\chi$($T$) of non-Curie-Weiss behavior is found in the
magnetization of Sm$^{+3}$ ions and ascribed to the comparable
size of $J$ multiplet to the $k{\rm_B}T$ in Sm$^{+3}$ ions Hund's
ground state of $J$ = 5/2 \cite{Morrish}.
Thus,  van Vleck contribution due to the higher level of
$J$ = 7/2 should be considered to account
for the observed susceptibility.
The observed magnetic susceptibility is described
according to the standard formula of
\be
\label{chi}
\chi(T) = N_{A} \left[ \frac{{\mu}^2_{\rm
eff}}{3k_{\rm B} ({T}-\Theta)} + \frac{20{\mu}^2_{\rm
B}}{7k_{\rm B}{\Delta}E} \right]
\ee
where the first term is a Curie-Weiss contribution
from the $J$ = 5/2 ground state multiplet,
and the second one is a temperature independent
van Vleck susceptibility due to coupling of the $J$ = 5/2 ground state
multiplet with the $J$ = 7/2 multiplet at an average energy
$k{\rm_B}{\Delta}E$ above ground state. The best fits for
the data of \Hp, \Hs, and powder average are plotted by solid
lines as shown in the Fig.~\ref{fig3}.
The fitting results are unsatisfactory for both field orientations
but apparently quite good for the powder-average case.
From the fitting results of the powder average data,
the splitting ${\Delta}E$, the effective moment $\mu_{\rm eff}$,
and the Curie-Weiss temperature $\Theta$ are extracted
to be 466 K, 0.36 $\mu_{\rm B}$, and $-$6.4 K, respectively.
The value of ${\Delta}E$ is smaller than
the of \psm, ($\approx$ 1150 K) \cite{Seaman}, which is
probably due to the doping of electrons by Ce$^{+4}$ ion and the
interaction between the localized and the doped electrons.

A particularly interesting feature in Fig.~\ref{fig3} is
that the susceptibility for \Hp\ reaches a minimum
and then increases slowly as the temperature increases still further,
which is similar to the one in Sm due to the small interval
between $J$ = 5/2 and $J$ = 7/2 multiplets.
This minimum susceptibility behavior is not observed for \Hs\
and powder averaged one,
which show monotonic decrease with increasing temperature.
One of the possible scenarios for this remarkable anisotropy is
that the splitting of $J$ multiplets has angular dependence.
It is conjectured that this can be caused by the non-negligible
anisotropic hybridization of conduction electrons with the localized
Sm$^{+3}$ ions and its angular dependence.
The heavy electronic behavior of \psm\ compound is manifested
itself by the relatively large $\gamma$
($\approx$ 82 mJ/mole${\cdot}$Sm${\cdot}$K$^2$) value,
which is the electronic specific contribution.
The large value of $\gamma$ is also found in the electron doped \scco\
in this study in specific heat measurements (see below).

\begin{figure}[t]
 \centerline{\psfig{figure=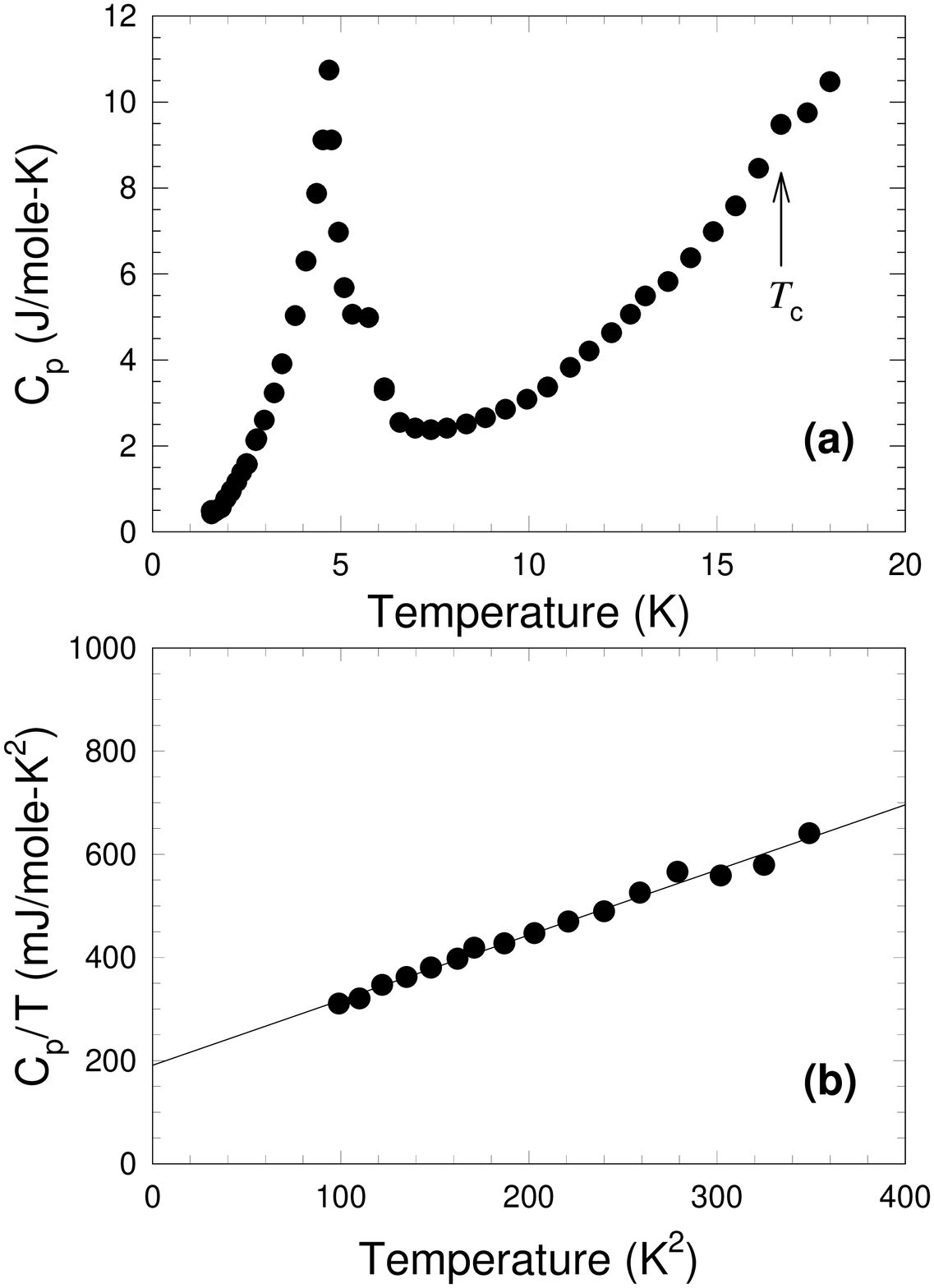,width=7cm,clip}}
 \caption{(a) Specific heat \cp\ versus temperature $T$ 
  of single crystal \scco\
  for 1.5 K $\le T \le$ 19.0 K.  (b) \cp\ versus $T^2$.}
 \label{fig4}
\end{figure}

The temperature dependent specific heat, \cp($T$), for \scco\
is plotted in Fig.~\ref{fig4}(a).
Clear evidence of a phase transition in \scco\ is given
by the sharp peak at $T$ = 4.7 K
and the superconducting transition is seen near \Tc\ $\approx$ 16.5 K
as a slight jump of \cp\, which is consistent with the \Tc\
from low field magnetization.
The data at $T$ = 5.7 K, which is level off the measured data,
is not understood yet and probably sample dependent
(or measurement error).
It was shown that a phase transition in \psm\
at $T \approx$ 6 K is attributed to AFM transition
from the $M$($T$) and \cp($T$) measurements \cite{Ghamaty}.
Thus, the observed transition in \scco\ is also of AF nature
and the \TN\ is shifted to lower temperature
with charge carrier (electrons) being doped.

In order to separate the magnetic and nonmagnetic contribution to \cp,
the data for 10 K $\le T \le$ 18 K is fitted to the equation
\be
\label{cpnm}
C_{\rm p}^{\rm NM}(T) = {\gamma}{T} + {\beta}{T}^3,
\ee
where the linear and the cubic terms correspond to the electronic
and lattice contributions to the specific heat, respectively.
The \cp$^{\rm NM}$($T$) for 10 K $\le T \le$ 18 K from Fig.~\ref{fig4}(a)
is plotted again with \cp$^{\rm NM}$($T$)$/T$ versus $T^2$
in Fig.~\ref{fig4}(b) together with the fitting values (solid line),
which shows nice agreement between the data and Equation
(\ref{cpnm}).
It is found that $\gamma$ = 191.0 (7) mJ/mole$\cdot$K$^2$ and
$\beta$ = 1.3 (1) mJ/mole$\cdot$K$^4$,
yielding the Debye temperature $\Theta_{\rm D}$ $\approx$ 219 K
from the relation of
$\Theta_{\rm D}$ $\propto$ (n/$\beta$)$^{1/3}$,
where $n$ is the number of atoms in a formula unit.
Although the above equation for the specific heat is valid
for temperatures below $\Theta_{\rm D}$/50 in usual metal,
the equation quite often works well for temperatures
below $\Theta_{\rm D}$/10 within an error of few percent,
which is satisfied in our temperature range \cite{Cho}.

\begin{figure}[t]
 \centerline{\psfig{figure=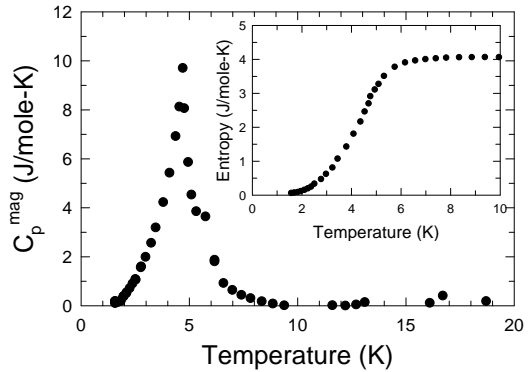,width=7cm,clip}}
 \caption{Magnetic specific heat (\cp$^{\rm mag}$)
   versus temperature $T$ of singel crystal \scco,
   \cp$^{\rm mag}$ = \cp\ $-$ \cp$^{\rm NM}$
   (= $\gamma$$T$+ $\beta$$T^3$)
   (see text).  Inset: entropy associated with the magnetic transition
   versus temperature.}
 \label{fig5}
\end{figure}

The observed value of $\gamma$ = 103.2 (7) mJ/mole$\cdot$Sm$\cdot$K$^2$
is significantly larger than those found in other \pln\ compounds,
0 $\pm$ 10 mJ/mole$\cdot$Nd$\cdot$K$^2$ for Ln = Nd,
1.3 $\pm$ 0.1 mJ/mole$\cdot$Pr$\cdot$K$^2$ for Ln = Pr \cite{Hundley}.
For \ncco\ compound,
which has highest superconducting transition temperature
among electron doped superconductors,
the value of $\gamma$ is enhanced to be
$\approx$ 29 mJ/mole$\cdot$Nd$\cdot$K$^2$ \cite{Ghamaty}.
It is natural to judge that the enhanced $\gamma$ is due to
the doped electrons.
Even for \psm, the $\gamma$ value was previously found to be exceptionally
large ($\approx$ 82 mJ/mole$\cdot$Sm$\cdot$K$^2$) \cite{Hundley}.
It was speculated that the effects of magnetic correlation exist well
above \TN\ $\approx$ 5.9 K,
thereby making accurate determination of $\gamma$ difficult.
However, our estimated $\gamma$ for \scco\ is still quite large
even though \TN\ is now lowered to be 4.7 K.
However, it is not clear now what is the origin of the large value of
$\gamma$ in the superconducting state.

The contribution of magnetic correlation to the measured \cp($T$)
is calculated as
\cp$^{\rm mag}$($T$) = \cp($T$) - \cp$^{\rm NM}$($T$),
where the extrapolation of \cp$^{\rm NM}$($T$) for low temperature
with the constants determined above is used, and is plotted
in Fig.~\ref{fig5}.
The entropy associated with the magnetic transition is calculated
from the \cp$^{\rm mag}$($T$) and its temperature dependence is plotted
in the inset of Fig.~\ref{fig5}.
The magnetic entropy saturates rapidly
above \TN\ to be $\approx$ 4.1 J/kmole,
indicating that the transition is driven by localized electrons.
However, the accumulated entropy is clearly smaller than 1.85$R$ln2,
where $R$ is gas constant,
which is the usual value of a doublet ground state of Sm$^{+3}$
\cite{Nekvasil}.
It was reported by \cite{Stewart2} that the magnetic entropy associated with
a magnetic transition is significantly reduced
if the magnetic transition is due to itinerant heavy fermionic
electrons in analogy to a BCS-type transition.
Thus, the reduced entropy can be explained by the fact that
itinerant electrons with heavy effective mass are involved
in the transition.
This explanation is also consistent with the anisotropic
temperature-dependent behavior of magnetization and
the enhanced electronic specific heat contribution.

\section{SUMMARY}

The single crystal of superconducting \scco\ compounds is studied
in terms of magnetization and specific heat measurements.
The largest difference in susceptibility, so far reported,
between \Hp\ and \Hs\ is found
and the temperature dependencies for both field orientations
do not follow the Curie-Weiss behavior
due to the small energy gap of $J$ = 7/2
multiplet above $J$ = 5/2 ground state.
With increasing temperature, the $\chi_{\parallel}$($T$) for \Hp\
exhibits a broad local minimum around $T$ = 100 K and a slow increase
while the  $\chi_{\perp}$($T$) for \Hs\
shows a monotonic decrease.
The specific heat data show a sharp peak at $T$ = 4.7 K,
which is of AF transition.
The estimated $\gamma$ value of electronic contribution is enhanced
with electron doping and clearly larger than the ones reported so far.
The entropy associated magnetic transition is obviously smaller than
the expected one of a doublet ground state.
Although the peculiar features found in this paper in \scco\ seem to
be related with the conduction electrons which is strongly interacting
with the localized electrons,
more study not only in experiments but also in theory should be done
to understand the magnetic and electronic properties and, further,
the mechanism of superconductivity in electron doped superconductors.

\acknowledgments We thank C. K. Kim for useful discussions and
communications. This work was supported by the Brain Korea 21
Project and by the Korea Science \& Engineering Foundation through
the grant No. 1999-2-114-005-5.

\end{multicols}

\end{document}